# Performance Gains of LLMs With Humans in a World of LLMs Versus Humans

Lucas McCullum[1,2] (0000-0001-9788-7987)
Pélagie Ami Agassi[3,*] (0000-0228-9295-1868)
Leo Anthony Celi[4,5,6] (0000-0001-6712-6626)
Daniel K Ebner[7] (0000-0001-6053-5865)
Chrystinne Oliveira Fernandes[4,5] (0000-0002-8623-9500)
Rachel S. Hicklen[8] (0000-0003-4317-5671)
Mkliwa Koumbia[9] (0009-0008-7900-1562)
Lisa Soleymani Lehmann[10] (0000-0001-8779-1244)
David Restrepo[11] (0000-0002-3789-1957)

*Corresponding Author (e-mail: agassiivy@gmail.com)
**Authors (except first author) listed in alphabetical order by last name

[1]Department of Radiation Oncology, The University of Texas MD Anderson Cancer Center, Houston, TX, USA
[2]UT MD Anderson Cancer Center UTHealth Houston Graduate School of Biomedical Sciences, Houston, TX, USA
[3]Department of Finance Bank Insurance, Centre de Formation Bancaire du Togo, Lomé, Togo
[4]Laboratory for Computational Physiology, Massachusetts Institute of Technology, Cambridge, MA, USA
[5]Department of Biostatistics, Harvard T.H. Chan School of Public Health, Boston, MA, USA
[6]Department of Medicine, Beth Israel Deaconess Medical Center, Boston, MA, USA
[7]Department of Radiation Oncology, Mayo Clinic, Rochester, MN, USA
[8]Research Medical Library, The University of Texas MD Anderson Cancer Center, Houston, TX, USA
[9]School of Engineering, The University of Tokyo, 2-11-16 Yayoi, Bunkyo-Ku, Tokyo 113-8656, Japan
[10]Department of Medicine, Mass General Brigham, Harvard Medical School, Boston, MA, USA
[11]CentraleSupélec, Université Paris-Saclay, Paris, France

**Funding Statement**

LM is supported by a National Institutes of Health (NIH) Diversity Supplement (R01CA257814-02S2). LAC is funded by the National Institute of Health through DS-I Africa U54 TW012043-01 and Bridge2AI OT2OD032701, the National Science Foundation through ITEST #2148451, and a grant of the Korea Health Technology R&D Project through the Korea Health Industry Development Institute (KHIDI), funded by the Ministry of Health & Welfare, Republic of Korea (grant number: RS-2024-00403047).

**Conflicts of Interest**

None.

**Acknowledgments**

The authors would like to acknowledge Dr. Danielle Bitterman for her thoughtful contributions to the early drafts of this work.

**Introduction**

    Since their introduction, transformer-based large language models (LLMs), as defined in a study by Wei et al. in 2022[1], have disrupted various fields including the field of medicine. Since the inception and public release of the most popular LLM, OpenAI's ChatGPT model, in November 2022, discovery that LLMs have passed the United States Medical Licensing Examination (USMLE) in February 2023[2], radiology board examinations in May 2023[3], and clinical reasoning examinations at Stanford in July 2023[4] has upended traditional beliefs. Currently, humans may be better at most tasks in medicine than LLMs, however several studies have shown well-defined text-based tasks where LLMs surpass human performance[5–7]. For example, a recent randomized clinical trial demonstrated that LLMs alone outperformed both a group of humans with and without access to LLMs in diagnostic reasoning applications[8], leading to the question of LLMs replacing humans on specific tasks. Currently, a considerable research effort is devoted to comparing LLMs to a group of human experts, where the term "expert" is often ill-defined or variable, at best, in a state of constantly updating LLM releases. Without proper safeguards in place, LLMs will threaten to cause harm to the established structure of safe delivery of patient care which has been carefully developed throughout history to keep the safety of the patient at the forefront. A key driver of LLM innovation is founded on community research efforts which, if continuing to operate under "humans versus LLMs" principles, will expedite this trend. Therefore, research efforts moving forward must focus on effectively characterizing the safe use of LLMs in clinical settings that persist across the rapid development of novel LLM models. In this communication, we demonstrate that rather than comparing LLMs to humans, there is a need to develop strategies enabling efficient work of humans with LLMs in an almost symbiotic manner.

    The problem with studies focused on comparing the output of LLMs with human experts is that these technologies move very quickly, with new OpenAI ChatGPT models being released approximately annually, for example, and community-developed models being released even faster. In a brief literature review of the field, we found 443 results when searching Pubmed for "ChatGPT-3.5" which has been discontinued after one year of use. We also found 64 results when searching Pubmed for "ChatGPT-3.5 Turbo" which has been discontinued after one year of use. Finally, we found 2 results when searching Pubmed for "ChatGPT o1-preview" which has been discontinued. Specifically, one of the studies identified directly compares ChatGPT o1-preview to GPT-4[9] and was published October 1, 2024, however ChatGPT o1-preview was discontinued just two months later when ChatGPT o1 was released on December 17, 2024. A further demonstration of the lifetime of some LLM models is shown in **Figure 1**.

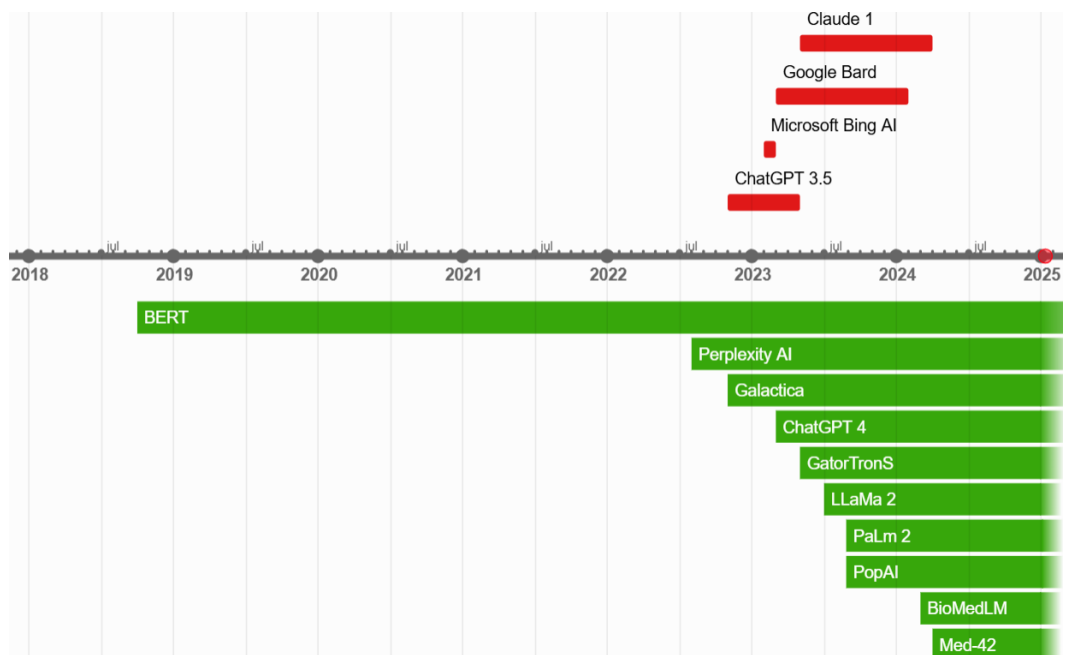

**Figure 1:** A timeline of the inception and discontinuation of the LLM models identified in our literature review of studies comparing LLM model output to human experts. The events above timeline in red represent LLM models which have been discontinued due to a newer version being released or a rebranding event (i.e., Google Bard rebranded to Gemini and Microsoft Bing AI rebranded to Copilot). The events below the timeline

in green represent LLM models which are still publicly available at the time of this writing. Over half (54%) of the models identified in our literature review that were compared to human experts fall into the red discontinued category representing the brevity of the utility that can be derived from these studies.

We conducted a broad literature review encompassing studies directly comparing LLM model performance to human experts using search terms developed in collaboration with a qualified medical librarian. For each study, the country of each author's institutional affiliation was determined using the Dimensions AI tool. A total of 1,567 studies were identified and compiled together to show a consistent increase in studies focusing on comparing LLMs to human experts as shown in **Figure 2**.

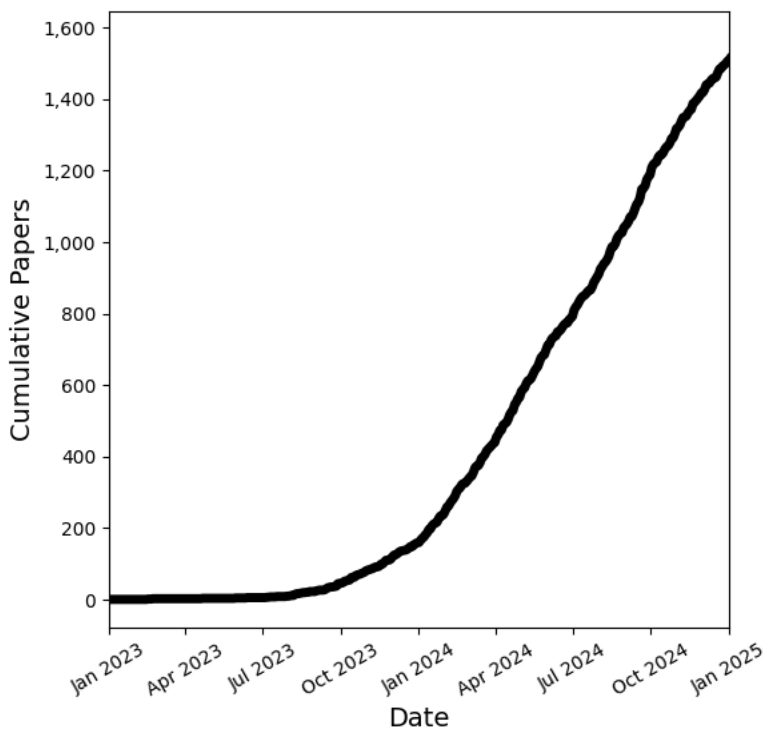

**Figure 2:** Exponential increase in the number of studies identified to compare LLMs to human experts.

To understand what kind of studies were being performed, we randomly selected 100 studies and found a total of 59 studies which compared LLMs to human experts after filtering. The most common field of comparison was in clinical decision making with 46% of studies followed by examination performance (27%), medical text writing (14%), and answering patient questions (14%). Of those, LLMs were compared to physicians the most at 59%, followed by non-physician medical professionals (15%), residents (12%), average test taker performance (10%), and medical students (3%). This highlights current discrepancy in what is considered to be a human expert leading to difficult interpretation of the clinical utility of the study results. More concerning is the size of the group of human experts used in the study with nearly 80% comparing LLMs to less than ten human experts as shown in **Figure 3**.

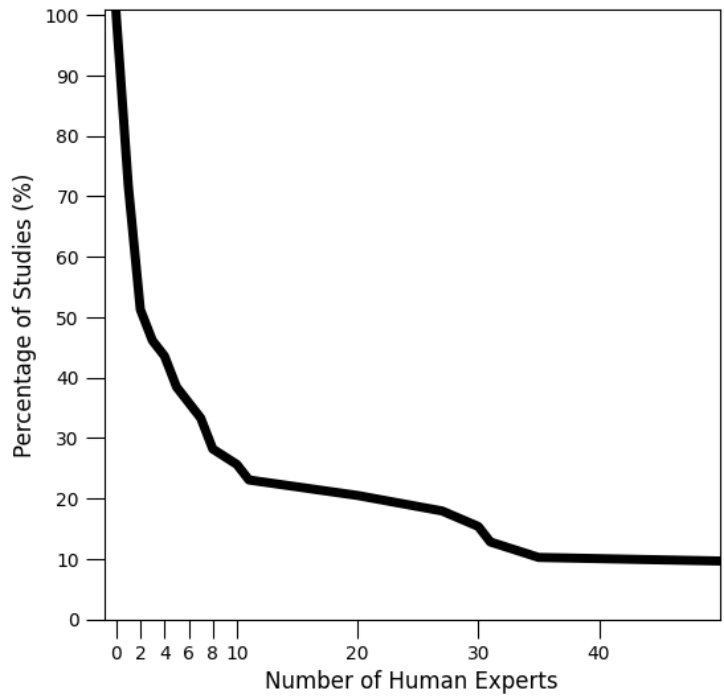

**Figure 3:** Proportional of studies as sorted by the number of human experts that they compare to with half of studies comparing to two or less human experts and nearly 80% of studies comparing to a cohort of less than ten human experts.

On top of this, comparing these models to human experts and board examinations is not sufficient if we want to know how they will behave in a real clinical setting where its accuracy becomes critical. Inevitably, the current model of medical academic research has an increasing lag-time behind these iteratively improved technologies. Even in the post-COVID era of necessitating rapid dissemination of literature through pre-print services like medRxiv and arXiv, peer-reviewed publishing times, which are the current standard for high quality science[10], are still on the order of months to years[11] which is too slow in the current era of rapid LLM generation and clinical translation to impact current trends. Papers showing human superiority may be incorrect once published; and those showing machine superiority may have repeated follow-up publications showing gradual improvement.

This process faces challenges on multiple fronts: both medical authors and reviewers at a medical journal, for instance, may be unaware of inherent contextual differences between LLM models (or context-awareness-simulating derivatives such as ChatGPT), contributing to a "blind leading the blind" effect[12]. As citations build, these blind spots are reinforced as potential truths in the literature, fed by an incomplete understanding of the technology. Comparison to human ability alone, however, does not further the field; it is just a current snapshot of the field. It particularly does not address how clinicians and patients may interact with current and future iterations of this technology to the best current technological potential. Therefore, a new strategy must be taken to publish effective LLM research. So what research is needed if we should stop comparing LLMs against human experts?

LLMs need to be treated in the same way as every medical device, tested with rigorous, randomized clinical trial, Level 1 studies designed with not only the endpoint of "Is this LLM better than the standard of human care?", but also "Is this LLM performing the same for all indicated patient demographics?", "Is it clear when this LLM is uncertain about a decision it is making[13–16]?", and "If approved, how should this LLM be deployed in the clinic to confidently, and safely, provide meaningful benefit to patients?". In practice, this will be difficult, however if we as a field wish to apply LLM models to the field of medicine which has an influence on patient outcomes then it should be treated with the same rigor as the currently existing medical devices such as drugs designed to treat disease, chemotherapy agents, and radiation therapy delivery devices. A beneficial first step would be a global database of LLM models that tracks their performance in the clinical setting across different populations in an open way without proprietary limitations. Without these interrogations in place, LLMs are simply blind content processors leading blind positions of authority given the current landscape of comparing the most recent

LLM iteration to human experts which quickly becomes defunct in the current environment of rapid growth of LLM solutions. Focusing on answers to the "how" questions in LLM applications in medicine is needed now more than the "what" questions. For example:

What is currently being researched in the LLM vs. human expert setting:
1. What is the performance of this specific, temporary, LLM on this single, untranslatable, task?
2. What questions can be asked to this specific, temporary, LLM that can be performed better by a small, unrepresentative, group of human experts?
3. What is the next area of medicine where I can apply the first two questions, conclude that "we should be cautious", before understanding how it works in this application?

Instead, what is needed in LLM research right now is:

1. How should this LLM be integrated into the current healthcare landscape?
2. Is the training data of the LLM sufficient to avoid data poisoning attacks[17]?
3. How well does this LLM perform in patient populations underrepresented in its training data and how does this compare to care delivered without LLMs?
4. How does knowing the answers to the "what" questions impact how this model is going to be applied to improved patient care and outcomes?
5. How do we compare LLMs to human performance in a robust way when needed[18]?
6. How should we develop ethical frameworks for LLM models[19]?
7. How do we integrate a true understanding of the world's rules[20] into model training and evaluation?
8. Effective reporting guidelines such as TRIPOD-LLM[21]
9. How do we address the large carbon footprint[22] of model training and execution at the mass-scale?
10. How do we address the inherent inequities between institutions due to the price of top models and compute required?
11. What regulations should be in place to protect patient privacy?
12. How do we address sycophancy bias?
13. How can we optimize the interactions between humans and LLMs?

**LLMs Lack Emotional Understanding: Collaborate with Human Experts in Emotional Intelligence**

In healthcare, emotion, also referred to as "emotional intelligence", plays an important role in patient outcomes, staff well-being, and overall healthcare system effectiveness[23]. The presence of compassion and empathy fosters trust, which in turn enhances communication and adherence to treatment protocols[24]. Leaders who possess emotional intelligence are adept at making well-rounded decisions that take into account both empirical data and ethical considerations, particularly in times of crisis. Furthermore, they cultivate a nurturing work environment that mitigates burnout and promotes staff retention. In addition, emphasizing emotional well-being leads to increased patient satisfaction and improved overall efficiency within healthcare systems. Ethical leadership is essential to ensure that financial limitations do not jeopardize patient care, while heightened emotional awareness bolsters resilience among healthcare personnel. By harmonizing emotional insight with logical reasoning, healthcare organizations can achieve a more compassionate and effective approach to care. Recent studies have shown that LLMs experience emotional intelligence which may benefit their deployment into healthcare settings[25–27], yet they still seem to lack general levels of empathy unless specifically guided[28]. However, LLMs are not sentient, which is dangerous for practice in healthcare when it comes to administering painkillers or anesthetists[29]. In fact, recognizing and alleviating pain is a fundamental ethical and medical responsibility, ensuring that patients do not suffer unnecessarily during treatment[30]. This inherent limitation of LLMs is where human intervention and collaboration is critical instead of comparing LLMs to human experts as if they will never interact with each other in the healthcare setting.

**LLMs Lack Local Knowledge: Bridge the Gap with Local Healthcare Experts**

Most LLMs are trained in English and primarily researched in English speaking countries as quantified from the "humans versus LLM" studies we identified and as shown in **Figure 4**. Therefore, as a direct consequence, they may not be accurate when used in other language settings leading to possible medical errors and misinterpretations[31]. Therefore, significant resources should be dedicated towards extending the language capabilities of LLMs due to the current heavy skew towards English speaking nations in training data and knowledge base[32,33]. This heavy skew will lead to the exaggeration of in-group and out-of-group bias already present in the current LLM models[34] that may lead to medical errors when being deployed in settings where training data is not represented. In addition, current proliferation of subscription-based payment plans for improved model performance will increase health inequalities in resource-poor nations. Therefore, the solution lies in the direct collaboration of LLMs with local human healthcare experts who can fill the gaps as the extra layer of intelligence as suggested by transfer learning. Why can't humans become this final layer?

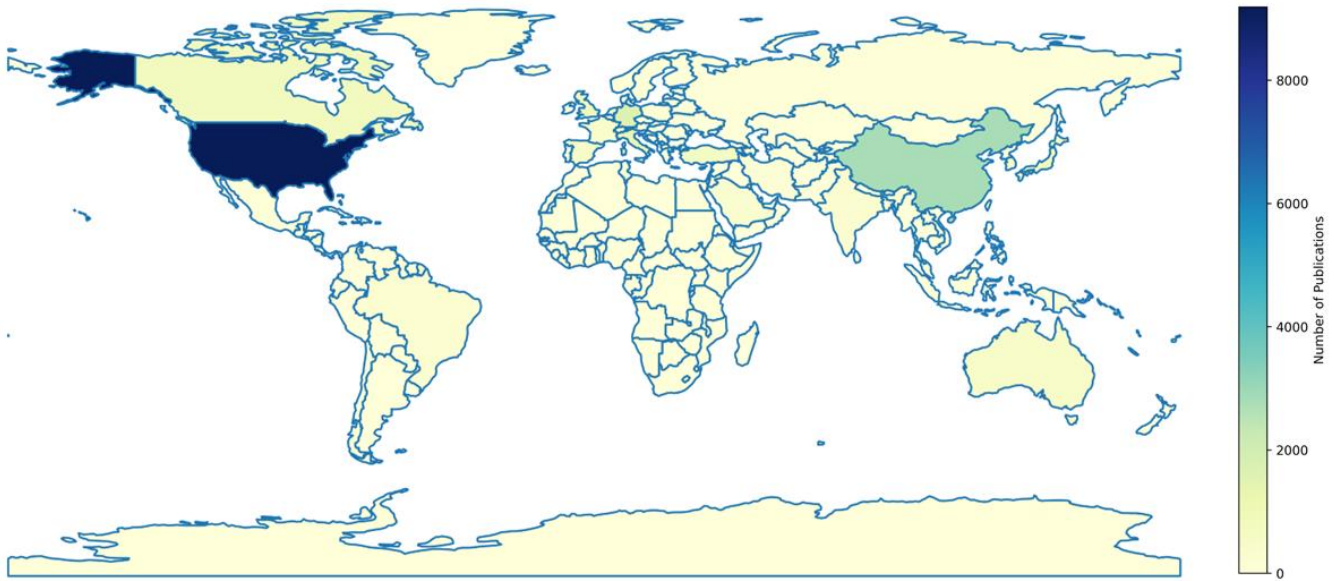

**Figure 4:** Global distribution of all author's country of affiliated institution from our identified studies conducting "humans versus LLMs" research. A vast majority of authors are from the United States followed by China and several minor contributing countries. A notable lack of representation is seen in low-and-middle income nations such as in the continent of Africa.

**Bridging the Gap for Enhanced Performance**

LLMs have undoubtedly demonstrated advanced performances that can help to improve and expedite many tasks, even in the medical sector. However, their extended usage as human replacement can cause harm to the established structure of safe delivery of patient care in medicine which has been carefully developed throughout history to keep the safety of the patient at the forefront. Therefore, research efforts moving forward must focus on effectively characterizing the safe use of LLMs in clinical settings that persist across the rapid development of novel LLM models.

Recent investigations have shown that LLMs struggle with integrating contextual features into its predictions[35] which has been enhanced through more advanced model architectures[36]. Such context can be enhanced through the intervention of human input to weight attention for the specific tasks needed to fine-tune a foundational LLM model[37]. The augmentation of LLM output with human input instead of competing against and replacing human input is how the largest performance gains from LLMs can be realized[38] which corroborates the already growing trend to increased collaboration between humans as a result of recent LLM releases[39]. Increased human-to-human collaboration in addition to human-to-LLM collaboration is the future of village mentoring and hive learning[40].

**Conclusion**

In summary, comparing LLMs to human experts needs to be limited to situations where its comparison is necessary for improved clinical outcomes and no other options exist. In this case, it should be conducted following existing guidelines (i.e., QUEST[18]) to uphold rigorous scientific study design such as when directed towards robust, longitudinal, safe implementation of these models in the clinical setting in an unbiased manner. As much as possible, research on LLMs in medicine should focus on maintaining strict adherence to model iteration agnostic conclusions, future-oriented intentions, and patient-centered ramifications.

LLMs are performing well and so are humans. However, both parties still have their strengths and weaknesses. Purely replacing humans with LLMs could cause harm to the medical society and patients as a direct result as LLMs are still evolving, are not sentient, and can not do some tasks that humans can do. Likewise, there are things that LLMs not only do better, but also faster than humans. For a better future, it is in the interest of the medical community to develop strategies enabling and catalyzing the collaboration between humans and LLMs so that we can move from "LLMs versus human experts" studies to "LLMs with human experts" studies.